\documentclass[11pt,showpacs, preprintnumbers,superscriptaddress,amsmath,amssymb,nofootinbib]{revtex4}
\usepackage{graphicx}
\usepackage{dcolumn}
\usepackage{bm}
\usepackage{amssymb}
\usepackage{amsmath}
\usepackage{epsfig}    
\usepackage{color}
\usepackage{slashed}
\usepackage{hhline}
\def\be{\begin{equation}}
\def\ee{\end{equation}}
\newcommand{\bea}{\begin{eqnarray}}
\newcommand{\eea}{\end{eqnarray}}
\newcommand{\nn}{\nonumber}

\numberwithin{equation}{section}
\usepackage{hyperref}

\begin{document}


\title{Neutrino mass model at a three-loop level from a non-holomorphic modular $A_4$ symmetry}

\author{Takaaki Nomura}
\email{nomura@scu.edu.cn}
\affiliation{College of Physics, Sichuan University, Chengdu 610065, China}

\author{Hiroshi Okada}
\email{hiroshi3okada@htu.edu.cn}
\affiliation{Department of Physics, Henan Normal University, Xinxiang 453007, China}

\date{\today}

\begin{abstract}
We study  a three-loop induced neutrino mass scenario from a non-holomorphic modular $A_4$ flavor symmetry and reach the minimum scenario leading predictions of the lepton masses, mixing angles, and Dirac and Majorana phases, which are shown through the chi square analyses. In addition, we discuss the lepton flavor violations, muon anomalous magnetic moment, lepton universality, and relic density of the dark matter candidate. And, we find that we need to extend our model if we satisfy the observed relic density of dark matter within the limit of perturbation where it can be done by adding one singlet scalar boson without changing predictions in neutrino sector.
 \end{abstract}

\maketitle

\newpage

\section{Introduction}
Thanks to a successful construction of non-holomorphic modular symmetry framework via Qu and Ding~\cite{Qu:2024rns},
we can safely deal with a beyond the standard model (BSM) without super-symmetric theories in using the framework for a flavor symmetry.
In fact, the non-holomorphic symmetries have been applied to some non-supersymmetric models~\cite{Ding:2024inn, Li:2024svh, Nomura:2024atp, Nomura:2024vzw, Nomura:2024nwh,Qu:2025ddz, Okada:2025jjo, Kobayashi:2025hnc, Loualidi:2025tgw, Nomura:2025ovm, Abbas:2025nlv}
in order to restrict the number of model parameters.
{In constructing a model, we have some advantage of applying non-supersymmetric framework to reduce number of new fields where extra fields are sometimes required to cancel a gauge anomaly in supersymmetric case.}

Radiatively induced neutrino mass models are one of the representative scenarios that do not demand the super-symmetric framework and one can naturally connect new particles to the SM model particles.
Sometimes, the model can possess the dark matter candidate(DM)~\cite{Ma:2006km} that often requests an additional symmetry to stabilize it.
Thus, constructing radiative neutrino mass models (with DM) using the non-holomorphic modular symmetry would be natural manner to make 
a model more attractive realizing more predictability. 


In our paper, we apply a non-holomorphic $A_4$ flavor symmetry to a well-known three loop neutrino mass model~\cite{Krauss:2002px}.
The three-loop neutrino model is phenomenologically interesting since the scale of new particles would be smaller compared to lower loop (or tree) level models due to loop suppression. 
 We then expect rich phenomenology such as collider and lepton flavor physics. 
Non-holomorphic modular symmetry framework is suitable to construct such a three loop model in minimal way; if we consider holomorphic one we need to add more fields to cancel gauge anomaly. 
Then we try to find the assignment under non-holomorphic modular $A_4$ symmetry as possible as minimal number of free parameters to fit the observables in lepton sector.
Through the chi square numerical analysis, we successfully search for the minimum model to predict the lepton masses and mixing angles in addition to reproduce the current neutrino observables in Nufit 6.0~\cite{Esteban:2024eli}.
Then, we perform further numerical analyses in order to satisfy lepton flavor violations (LFVs), muon anomalous magnetic moment, (muon $g-2$), lepton universality, and the dark matter (DM). 
{As a result, we find that relic density is too large within the limit of perturbation requiring a new interaction where it can be done by adding one singlet scalar boson without changing predictions in neutrino sector.}

This paper is organized as follows.
In Sec.~II, we explain our minimum three-loop neutrino mass model constructing the renormalizable Lagrangian in the lepton sector, Higgs sector, the charged-lepton sector, heavier Majorana fermion sector,  and the active neutrino sector.
Then, we formulate the LFVs, muon $g-2$, lepton universality, and relic density of the DM.
In Sec.~III, we perform $\chi$ square analysis and show some predictions for normal and inverted hierarchies in the neutrino sector. Employing the benchmark points of the best fit values in the lepton sector, we further demonstrate the numerical analyses 
for the LFVs, muon $g-2$, lepton universality, and the relic density of DM. 
We have conclusions and discussion in Sec.~IV.
In Appendix A, we show the three-loop function in the neutrino sector.

\begin{center} 
\begin{table}[tbh!]
\begin{tabular}{|c||c|c|c||c|c|c|}\hline\hline  
  & \multicolumn{3}{c|}{Leptons} & \multicolumn{3}{c|}{Bosons}   \\ \hline \hline
& ~$ \overline{L_L}$~& ~${\ell}_R$~& ~$N_R$~& ~$H$~& ~$S^+_1$~& ~$S^+_2$~       \\\hline\hline 
$SU(2)_L$ & $\bm{2}$  & $\bm{1}$& $\bm{1}$   & $\bm{2}$ & $\bm{1}$ & $\bm{1}$   \\\hline 
$U(1)_Y$   & $-\frac12$ & $1$ & $0$ & $\frac12$   & $+1$ & $+1$    \\\hline
 $A_4$ & $3$ & $\{ 1\}$ & $3$ & $1$ & $1$  & $1$         \\ \hline
$-k_I$ & $-1$ & $+1$& $0$ & $0$ & $+2$& $-1$    \\
\hline
\end{tabular}
\caption{Field contents and  their charge assignments in the model under $SU(2)_L\times U(1)_Y\times A_4$ where $-k_I$ is the number of modular weight.  Here $\{1\}$ stands for the combination of $A_4$ singlets $\{1,1',1''\}$ .}
\label{tab:1}
\end{table}
\end{center}

 \section{Model setup}
{In this section, we show setup of the model based on $G_{\rm SM} \times A_4$ symmetry where $G_{\rm SM}$ being the SM gauge symmetry and $A_4$ is modular one. In lepton sector, We introduce singlet fermion which is triplet under $A_4$ with modular weight 0. In scalar sector, we introduce two charged singlets distinguished by modular weight $+2$ and $-1$. The SM leptons $\overline{L_L}$ and $\ell_R$ are also $A_4$ triplets with modular weight $-1$ and $+1$ respectively. The assignments are summarized in Table~\ref{tab:1}. By the assignments of modular weight, we can eliminate unwanted terms like $\overline{N_R} L_L H$ and neutrino masses are generated at three-loop level as discussed below.
}

The relevant Lagrangian under these symmetries is given by  
\begin{align}
 - {\cal L}_\ell & = 
{a_e} \left[y_1 \overline{L_{L_e}} +y_2 \overline{L_{L_\tau}}+y_3 \overline{L_{L_\mu}}\right] e_R H
+
{a_\mu} \left[y_2 \overline{L_{L_\mu}} +y_3 \overline{L_{L_e}}+y_1 \overline{L_{L_\tau}}\right] \mu_R H
 \nn \\
& + {a_\tau} \left[y_3 \overline{L_{L_\tau}} +y_1 \overline{L_{L_\mu}}+y_2 \overline{L_{L_e}}\right] \tau_R H
\nn\\
& +{a_\nu} 
\left[y_1( \overline{L_{L_\mu}} \cdot {L^C_{L_\tau}} - \overline{L_{L_\tau}} \cdot {L^C_{L_\mu}})
+y_2( \overline{L_{L_\tau}} \cdot {L^C_{L_e}} - \overline{L_{L_e}} \cdot {L^C_{L_\tau}})
+y_3( \overline{L_{L_e}} \cdot {L^C_{L_\mu}} - \overline{L_{L_\mu}} \cdot {L^C_{L_e}})
\right] S^-_1
\nn\\
&  + {b_\nu} \overline{e^C_R} \left[y_1 N_{R_1} +y_2 N_{R_3}+y_3 N_{R_2} \right]  S^+_2
+
{c_\nu} \overline{\mu^C_R} \left[y_2 N_{R_2} +y_3 N_{R_1}+y_1 N_{R_3 }\right] S^+_2 
 \nn \\
&  + {d_\nu} \overline{\tau^C_R} \left[y_3 N_{R_3} +y_1 N_{R_2}+y_2 N_{R_1} \right]  S^+_2   \nn\\
&+M_1(\overline{N^C_{R_1}}N_{R_1}+\overline{N^C_{R_2}}N_{R_3}+\overline{N^C_{R_3}}N_{R_2})
\nn\\
&M_2 
\left[y_1 (2\overline{N^C_{R_1}}N_{R_1}-\overline{N^C_{R_2}}N_{R_3}-\overline{N^C_{R_3}}N_{R_2} )
+y_2 (2\overline{N^C_{R_2}}N_{R_2}-\overline{N^C_{R_1}}N_{R_3}-\overline{N^C_{R_3}}N_{R_1} )\right.\nn\\
&\left. +y_3 (2\overline{N^C_{R_3}}N_{R_3}-\overline{N^C_{R_1}}N_{R_2}-\overline{N^C_{R_2}}N_{R_1} )
\right]
 +{\rm h.c.}, 
\label{yukawa}
\end{align}
where we define $Y_3^{(0)} = [y_1,y_2,y_3]$~\cite{Qu:2024rns} and $"\cdot"$ indicates $i \sigma_2$ factor to make the term $SU(2)_L$ invariant. 
The first two terms generates the mass of charged-leptons and parameters $\{a_e,a_\mu,a_\tau \}$ are real without loss of generality by rephasing them into $e_R,\mu_R,\tau_R$, respectively.

{
\subsection{Scalar sector}
The scalar potential in the model is given by
\begin{align}
  {\cal V} &= \mu_H^2 |H|^2 + \mu^2_{S_1} |S_1^+|^2+ \mu^2_{S_2} |S_2^+|^2
  + \lambda_0 [ (S_1^+ S_2^-)^2 +{\rm h.c.} ] \nn\\  
&+ \lambda_H |H|^4 + \lambda_{S_1} |S_1^+|^4 + \lambda_{S_2} |S_2^+|^4 
+ \lambda_{HS_1} |H|^2|S_1^+|^2 + \lambda_{HS_2} |H|^2|S_2^+|^2
+ \lambda_{S_1S_2} |S_1^+|^2 |S_2^+|^2
  .\label{Eq:pot}
\end{align}
%
The SM Higgs field is denoted by
\begin{equation}
H=
\begin{pmatrix} w^+ \\ \frac{v + \tilde{h}+i z }{\sqrt2} \end{pmatrix} ,
\end{equation} 
and $v\approx 246$ GeV is  vacuum expectation value (VEV) in the Higgs basis after the spontaneous symmetry breaking, $z$ is absorbed by the neutral gauge boson of the SM $Z$, and  $w^+$ is absorbed by the charged gauge boson of the SM $W^+$.
{The charged scalar masses are respectively given by
\begin{align}
& m_{S_1}^2 = \mu^2_{S_1} + \frac12 \lambda_{H S_1} v^2, \\ 
& m_{S_2}^2 = \mu^2_{S_2} + \frac12 \lambda_{H S_2} v^2. 
\end{align}
In the numerical analysis we consider $m_{S_{1,2}}$ to be free parameters.}

\subsection{Charged-lepton mass matrix}
After the spontaneous electroweak symmetry breaking,
the charged-lepton mass matrix $M_e$ is given by
\begin{align}
&M_e = \frac{v}{\sqrt2}
\begin{pmatrix}
 y_1 &  y_3 & y_2 \\ 
 y_3 &  y_2& y_1 \\ 
y_2 &   y_1 & y_3 \\ 
\end{pmatrix}
\begin{pmatrix}
 a_e &  0 & 0 \\ 
0 & a_\mu & 0 \\ 
0 & 0 & a_\tau \\ 
\end{pmatrix} .
 \label{massmat}
\end{align}
Then, the charged-lepton mass matrix is diagonalized by a bi-unitary mixing matrix as $D_{\ell} \equiv{\rm diag}(m_e,m_\mu,m_\tau)=V^\dag_{eL} M_e V_{eR}$. Therefore, $\ell_{L(R)}\equiv V_{eL(R)} \ell'_{L(R)}$
where $ \ell'_{L(R)}$ is the mass eigenstate.
These three parameters are used in order to fit the mass eigenvalues of charged-leptons by solving the following three relations:
\begin{align}
&{\rm Tr}[M_e M_e^\dag] = |m_e|^2 + |m_\mu|^2 + |m_\tau|^2,\\
&{\rm Det}[M_eM_e^\dag] = |m_e|^2  |m_\mu|^2  |m_\tau|^2,\\
&({\rm Tr}[M_eM_e ^\dag])^2 -{\rm Tr}[(M_e M_e^{\dag})^2] =2( |m_e|^2  |m_\mu|^2 + |m_\mu|^2  |m_\tau|^2+ |m_e|^2  |m_\tau|^2 ).
\end{align}
For our convenience to construct the neutrino mass matrix below, we define $\tilde D_\ell$ that is given by $D_\ell\equiv m_\tau \tilde D_\ell$.

\subsection{Heavier Majorana fermion mass matrix}
The heavier Majorana mass matrix is given by
\begin{align}
&M_N = M_1
\left[
\begin{pmatrix}
 1 &  0 & 0 \\ 
0 & 0 & 1 \\ 
0 & 1 & 0 \\ 
\end{pmatrix} 
+ \tilde M_2
\begin{pmatrix}
2 y_1 & - y_3 & -y_2 \\ 
- y_3 &  2y_2& -y_1 \\ 
-y_2 & -  y_1 & 2 y_3 \\ 
\end{pmatrix}
\right]
\equiv M_1 \tilde M_N ,
 \label{massmat}
\end{align}
where $\tilde M_2\equiv M_2/M_1$ can be real without loss of generality.
$M_N$ is diagonalized by $D_N\equiv U_N^T M_N U_N$ ($\tilde D_N \equiv  U_N^T \tilde M_N U_N$)., therefore
$N_R\equiv U_N \psi_R$. Here, $\psi_R$ is the mass eigenstate.

\subsection{Active neutrino mass matrix} 
\label{neut}
The active neutrino mass matrix is given at three-loop level via the following Lagrangian 
in terms of mass eigenstates 
\begin{align}
a_\nu \left( \overline{\nu_L} H \ell'^C_{L} + \overline{\ell'_L} H^T \nu^C_{L}\right) S^-_1
+b_\nu \overline{\ell'^C_R} Y \psi_R  S^+_2 
 +{\rm h.c.},
\label{yukawa}
\end{align}
where $H\equiv h V^*_{eL}$ and $Y\equiv V^T_{eR} y U_N$.
The Yukawa matrices $y$ and $h$ are respectively found as
\begin{align}
h& = 
\begin{pmatrix}
 0 &  y_3 &- y_2 \\ 
- y_3 & 0 & y_1 \\ 
y_2 & - y_1 & 0 \\ 
\end{pmatrix},\\
y & = 
\begin{pmatrix}
 1 &  0 & 0 \\ 
0 & \tilde c_\nu & 0 \\ 
0 & 0 &\tilde d_\nu \\ 
\end{pmatrix}
\begin{pmatrix}
 y_1 & y_3 & y_2 \\ 
 y_3 & y_2& y_1 \\ 
y_2 &  y_1 & y_3 \\ 
\end{pmatrix} , 
\end{align}
where $\tilde c(\tilde d)_\nu\equiv c(d)_\nu/b_\nu$ are complex free parameters.
The neutrino mass matrix is then given by
\begin{align}
 (m_{\nu})_{ij}  &\approx - \frac{\lambda_0 (a_\nu b_\nu)^2}{(4 \pi)^6} \frac{m^2_\tau}{M_1}
H^* \tilde D_\ell Y^* \tilde D_{N} F Y^\dag \tilde D_\ell H^\dag \equiv \kappa \tilde m_\nu,
%
%
\end{align}
Here, $F$ is a loop function via three loop diagram and it depends on the mass eigenvalues of $\{\psi_R, S_1^+, S_2^+\}$.~\footnote{In general, the loop function also depends on the masses of charged-leptons. However, we suppose these masses to be negligibly tiny compared to the exotic particles inside the loop.}
 Since the masses of $\psi_R$ contribute to the structure of neutrino mass matrix, there would be too many 
 free parameters to get some predictions for the neutrino sector.
Thus, we consider a special situation among the mass hierarchies  of $\psi_R, S_1^+, S_2^+$ so that 
$F$ is independent of the structure of neutrino mass matrix.
When we assume $D_{N_i}
 \ll  m_{S_1}\sim m_{S_2}$, one finds that the dominant part of the loop-function $F$ is a constant and can explicitly be given by  $F\approx 0.062$. In detail, one finds Appendix A.
Thus,  we redefine the neutrino mass matrix as follows:
\begin{align}
\kappa&\equiv  - \frac{\lambda_0 F (a_\nu b_\nu)^2}{(4 \pi)^6} \frac{m^2_\tau}{M_1}, \label{eq:kappa}\\
\tilde m_\nu &\equiv H^* \tilde D_\ell Y^* \tilde D_N Y^\dag \tilde D_\ell H^\dag.
\end{align}
%
The dimensionless matrix $\tilde m_\nu$ is diagonalized by a unitary matrix $U_\nu$ as $U_\nu^T \tilde m_\nu U_\nu =\tilde D_\nu$,
where  $\tilde D_\nu = {\rm diag}[\tilde D_{\nu_1},\tilde D_{\nu_2},\tilde D_{\nu_3}]$
and the Pontecorvo-Maki-Nakagawa-Sakata unitary matrix $U_{\rm PMNS}$ is defined by $V_{eL}^\dag U_\nu$. 
Note here that the lightest neutrino mass is zero due to two matrix rank of the neutrino.  
The atmospheric mass squared difference $\Delta m^2_{\rm atm}$ is thus found as 
\begin{align}
&{\rm NH}:\ \Delta m^2_{atm}= \kappa^2 \tilde D^2_{\nu_3} ,\\
&{\rm IH}:\ \Delta m^2_{atm}= \kappa^2 \tilde D^2_{\nu_2} ,
\end{align}
where NH(IH) represents normal(inverted) hierarchy.
The solar mass squared difference $\Delta m^2_{\rm sol}$ is given by
\begin{align}
&{\rm NH}:\Delta m^2_{\rm sol}= \kappa^2 \tilde D^2_{\nu_2},
\\
&{\rm IH}:\ \Delta m^2_{\rm sol}= \kappa^2 (\tilde D^2_{\nu_2} - \tilde D^2_{\nu_1} ).
\end{align}
The effective mass for neutrinoless double beta decay is given by
\begin{align}
&{\rm NH}: \langle m_{ee}\rangle = \kappa
 \left|
 +\tilde D_{\nu_2} s^2_{12} c^2_{13}e^{i\alpha_{21}}
+\tilde D_{\nu_3} s^2_{13}e^{-2i\delta_{CP}} \right|,\\
&{\rm IH}: \langle m_{ee}\rangle = \kappa \left|\tilde D_{\nu_1} c^2_{12} c^2_{13}+
\tilde D_{\nu_2} s^2_{12} c^2_{13}e^{i\alpha_{21}}
\right|,
\end{align}
where Majorana phase is defined by ${\rm diag}[1,e^{i\alpha_{21}/2} ,1]$ and we adopt the standard parametrization for the PMNS unitary matrix.
A current KamLAND-Zen data~\cite{KamLAND-Zen:2024eml}. provides measured observable in future and 
its upper bound is given by  $\langle m_{ee}\rangle<(28-122)$ meV at 90 \% confidence level.
%
The minimal cosmological model
$\Lambda$CDM $+\sum D_{\nu}$ provides upper bound on $\sum D_{\nu}\le$ 120 meV~\cite{Vagnozzi:2017ovm, Planck:2018vyg}.
Moreover, recently combination of DESI and CMB data gives more stringent upper bound on this bound;
$\sum D_{\nu}\le$ 72 meV~\cite{DESI:2024mwx}. 

\subsection{Lepton Flavor Violations and Muon Anomalous Magnetic Moment}
{\it $\ell_\alpha \to \ell_\beta \gamma$ process}: 
First of all, let us consider
the processes $\ell_\alpha \to \ell_\beta \gamma$ at one-loop
level~\footnote{The experimental bounds are summarized in
  Table~\ref{tab:Cif}.}.  The formula for the branching ratio can
generally be written as
\begin{align}
{\rm BR}(\ell_\alpha \to \ell_\beta \gamma)
=
\frac{48\pi^3 C_{\alpha\beta} \alpha_{\rm em}}{{\rm G_F^2} m_\alpha^2 }\,
  (|(a_R)_{\alpha \beta}|^2+|(a_L)_{\alpha \beta}|^2),
\end{align}
where $\alpha_{\rm em}\approx1/137$ is the fine-structure constant,
$C_{\alpha\beta} \approx(1,0.1784, 0.1736)$ for ($(\alpha,\beta)=(\mu,e),(\tau,e),(\tau,\mu)$), 
${\rm G_F}\approx1.17\times 10^{-5}$ GeV$^{-2}$ is the Fermi constant, 
and $a_{L/R}$ is respectively given by
\begin{align}
(a_{R})_{\alpha \beta}&\approx
\frac1{(4\pi)^2}\sum_{a=e,\mu,\tau}\sum_{i=1}^3
\left( a_\nu^2 \frac{H_{\beta i} H^\dag_{i \alpha} }{12 m^2_{S_1}} m_{\ell_\alpha} +
b_\nu^2
\frac{ Y^*_{\beta i} Y^T_{i \alpha}  }{m^2_{S_2}} m_{\ell_\beta} F_I\left[\frac{D_{N_i}^2}{m_{S_2}^2}\right]  \right),
\\
(a_{L})_{\alpha \beta}&=
\frac1{(4\pi)^2}\sum_{a=e,\mu,\tau} \sum_{i=1}^3
\left( a_\nu^2 \frac{ H_{\beta i} H^\dag_{i \alpha}  }{12 m^2_{S_1}} m_{\ell_\beta} +
b_\nu^2
\frac{Y^*_{\beta i} Y^T_{i \alpha}  }{m^2_{S_2}} m_{\ell_\alpha}  F_I\left[\frac{D_{N_i}^2}{m_{S_2}^2}\right]   \right),
 \end{align} 
where 
\begin{align}
&F_I(x)=
\frac{1-6x+3 x^2+2 x^3-6x^2\ln[x]}{6(1-x)^4}.
\end{align}
Once we assume that $m_{\ell_\alpha} \gg m_{\ell_\beta}$, the formula can be simplified to
\begin{align}
{\rm BR}(\ell_\alpha \to \ell_\beta \gamma)\approx
\frac{48\pi^3 C_{\alpha\beta} \alpha_{\rm em}}{{\rm G_F^2}(4\pi)^4 }
\left[
\frac{a_\nu^4}{ 144 m^4_{S_1}} \left|\sum_{a=e,\mu,\tau} H_{\beta a} H^\dag_{a \alpha}\right|^2 
+
\frac{b_\nu^4} {m^4_{S_2}}
\left|\sum_{i=1}^3 
Y^*_{\beta i} Y^T_{i \alpha}   F_I\left[\frac{D_{N_i}^2}{m_{S_2}^2}\right]
\right|^2  
\right].
\end{align}

 \begin{table}[t]
\begin{tabular}{c|c|c|c} \hline
Process & $(\alpha,\beta)$ & Experimental bounds ($90\%$ CL) & References \\ \hline
$\mu^{-} \to e^{-} \gamma$ & $(\mu,e)$ &
	${BR}(\mu \to e\gamma) < 4.2 \times 10^{-13}$ & \cite{MEG:2016leq} \\
$\tau^{-} \to e^{-} \gamma$ & $(\tau,e)$ &
	${BR}(\tau \to e\gamma) < 3.3 \times 10^{-8}$ & \cite{MEG:2013oxv} \\
$\tau^{-} \to \mu^{-} \gamma$ & $(\tau,\mu)$ &
	${BR}(\tau \to \mu\gamma) < 4.4 \times 10^{-8}$ & \cite{MEG:2013oxv}   \\ \hline
\end{tabular}
\caption{Summary for the experimental bounds of the LFV processes 
$\ell_\alpha \to \ell_\beta \gamma$.}
\label{tab:Cif}
\end{table}

\begin{table}[t]
\begin{tabular}{c|c|c} \hline
Process  & Experiments & Bound ($90\%$ CL)  \\ \hline
{\rm Lepton/hadron\ universality}  &
$\sum_{q=b,s,d}|V^{\rm exp}_{uq}|^2=0.9999\pm0.0006$:  & $|H^\dag_{e\mu}|^2<0.007\left(\frac{m_{S_1}}{a_\nu {\rm TeV}}\right)^2$   \\ 
${\rm \mu/e\ universality}$ & 
	$\frac{G_\mu^{\rm exp}}{G_e^{\rm exp}}=1.0010\pm0.0009$ & $||H^\dag_{\mu\tau}|^2-|H^\dag_{e\tau}|^2|<0.024\left(\frac{m_{S_1}}{a_\nu {\rm TeV}}\right)^2$  \\ 
${\rm \tau/\mu\ universality}$ & 
	$\frac{G_\tau^{\rm exp}}{G_\mu^{\rm exp}}=0.9998\pm0.0013$ & $||H^\dag_{e\tau}|^2-|H^\dag_{e\mu}|^2|<0.035\left(\frac{m_{S_1}}{a_\nu {\rm TeV}}\right)^2$  \\ 
${\rm \tau/e\ universality}$ & 
	$\frac{G_\tau^{\rm exp}}{G_e^{\rm exp}}=1.0034\pm0.0015$ & $||H^\dag_{\mu\tau}|^2-|H^\dag_{e\mu}|^2|<0.04\left(\frac{m_{S_1}}{a_\nu {\rm TeV}}\right)^2$  \\ \hline
\end{tabular}
\caption{Summary of the lepton universality and the corresponding bounds
on $f_{\alpha\beta}$.}
\label{tab:lep-univ}
\end{table}

The formula for the muon $g-2$ can be written in terms of $a_L$ and $a_R$, 
and simplified as follows:
\begin{align}
\Delta a_\mu\approx -{m_\mu}(a_R+a_L)_{\mu \mu}
\approx
-\frac{m^2_\mu}{(4\pi)^2}
\sum_{a=e,\mu,\tau} \sum_{i=1}^3
\left( a_\nu^2 \frac{H_{\mu a} H^\dag_{a \mu}  }{6 m^2_{S_1}} 
+ 2 b_\nu^2
\frac{Y^*_{\mu i} Y^T_{i \mu}  }{m^2_{S_2}}  F_I\left[\frac{D_{N_i}^2}{m_{S_2}^2}\right]   \right)
.\label{damu}
\end{align}
Notice here that this contribution to the muon $g-2$ is negative, yet it
is negligible compared to the deviation in the experimental 
value~${\cal O}(10^{-9})$~\cite{Muong-2:2006rrc}.

\subsection{ Lepton Universality}
Here, we just employ the results of lepton universality from precursor's works~\cite{Herrero-Garcia:2014hfa}
whose results provide us the upper bounds on coupling $H$ in terms of $m_{S_1}$ and $a_\nu$.
We summarize these results in Table~\ref{tab:lep-univ}.

\subsection{Dark Matter}
{\it Relic density}:
Our DM is identified as  the lightest Majorana fermion $N_1$ where we denote $N_1$ as $X$ hereafter and its mass is $m_\chi$.
In order to analyze it simpler, 
we impose the following condition,
$1.2 m_\chi\lesssim D_{N_2}\le D_{N_3}$, in order to evade an effect of co-annihilation interactions for the relic density of DM.
\footnote{More detailed computations are found in \cite{Ahriche:2013zwa, Cheung:2016ypw}.}
Under the condition, the dominant contribution to the relic density arises from $Y$.
 Then, the non-relativistic cross section is expanded by relative velocity $v_{\rm rel}^2$; $(\sigma v_{\rm rel})\approx a_{\rm eff}
 + b_{\rm eff} v^2_{\rm rel}+ {\cal O}(v^4_{\rm rel})$ and  found as follows:
\begin{align}
(\sigma v_{\rm rel})\approx
\frac{ m_\chi^2}{48\pi (m^2_{S_2} + m^2_\chi)^4}
\left(
m_{S_2}^2 + 2 m_{S_2}^2 m_\chi^2 + 3 m_\chi^4
\right)  b_\nu^4 
 \sum_{a,b=1}^3 |Y^*_{ai} Y^T_{1,b}|^2
v_{\rm rel}^2 ,  \label{eq:DMCX}
\end{align}
where we have neglected the masses of charged-leptons.
The above cross section suggests that it is p-wave dominant. 
The relic density is then given by
\begin{align}
\Omega h^2\approx 
\frac{1.07\times10^9 }{\rm GeV}
\frac{x_f^2}{3 \sqrt{g^*} M_P b_{\rm eff} 
},
\end{align}
where $g^*\approx100$, $M_P\approx 1.22\times 10^{19}{\rm GeV}$, $x_f\approx20$.
In our numerical analysis below, we use
rather relaxed experimental range $0.11\le \Omega h^2\le 0.13$,
since we simplify our analysis of the relic density. 

\section{Numerical analysis}

In this section, we demonstrate numerical analyses in light of all the experimental results which we discussed before.
Then, we show the results of LFVs, lepton $g-2$, and DM.

\subsection{Numerical results of the lepton sector}

At first, we perform $\chi$ square analysis adopting data from NuFit6.0~\cite{Esteban:2024eli},
where we have adopt five reliable observables; three mixings, two mass square differences, for the analysis. 
The yellow points represents the interval of $2\sigma-3\sigma$, and red one $3\sigma-5\sigma$,
where we would not find any solutions within $2\sigma$.
Our three input parameters randomly select within the following range:
\begin{align}
&\{\tilde M_2, |\tilde c_\nu|, |\tilde d_\nu| \}  \in [10^{-5},10^5],
\end{align}
where we work on the fundamental region of $\tau$ and $\tilde c_\nu, \tilde d_\nu$ are complex.

 After the numerical analysis, we find that IH case is not favored in the model where minimal $\chi^2$ can be at most  $\mathcal{O}(1500)$. 
We thus summarize our results regarding only NH case in next subsection.
Note that the parameters $\{a_e, a_\mu, a_\tau \}$ are chosen to fit the observed charged lepton masses and $\{a_\nu, b_\nu, M_1\}$ are related to fix the scale of neutrino mass via $\kappa$ defined in Eq.~\eqref{eq:kappa}.
Thus relative neutrino mass and three mixing angles are fitted using remaining parameters $\{\tau, c_\nu, d_\nu, M_2\}$ corresponding to 7 real parameters. 
Since three of these real parameters are related to complex phases it is not trivial if we can fit the neutrino data. 
In fact, we would not be able to obtain any solutions in IH case.
To improve the fitting further such as IH, we need to change assignment of modular weight to increase the number of free parameter.

\subsection{Neutrino observables in NH case}


\begin{figure}[t]
  \includegraphics[width=77mm]{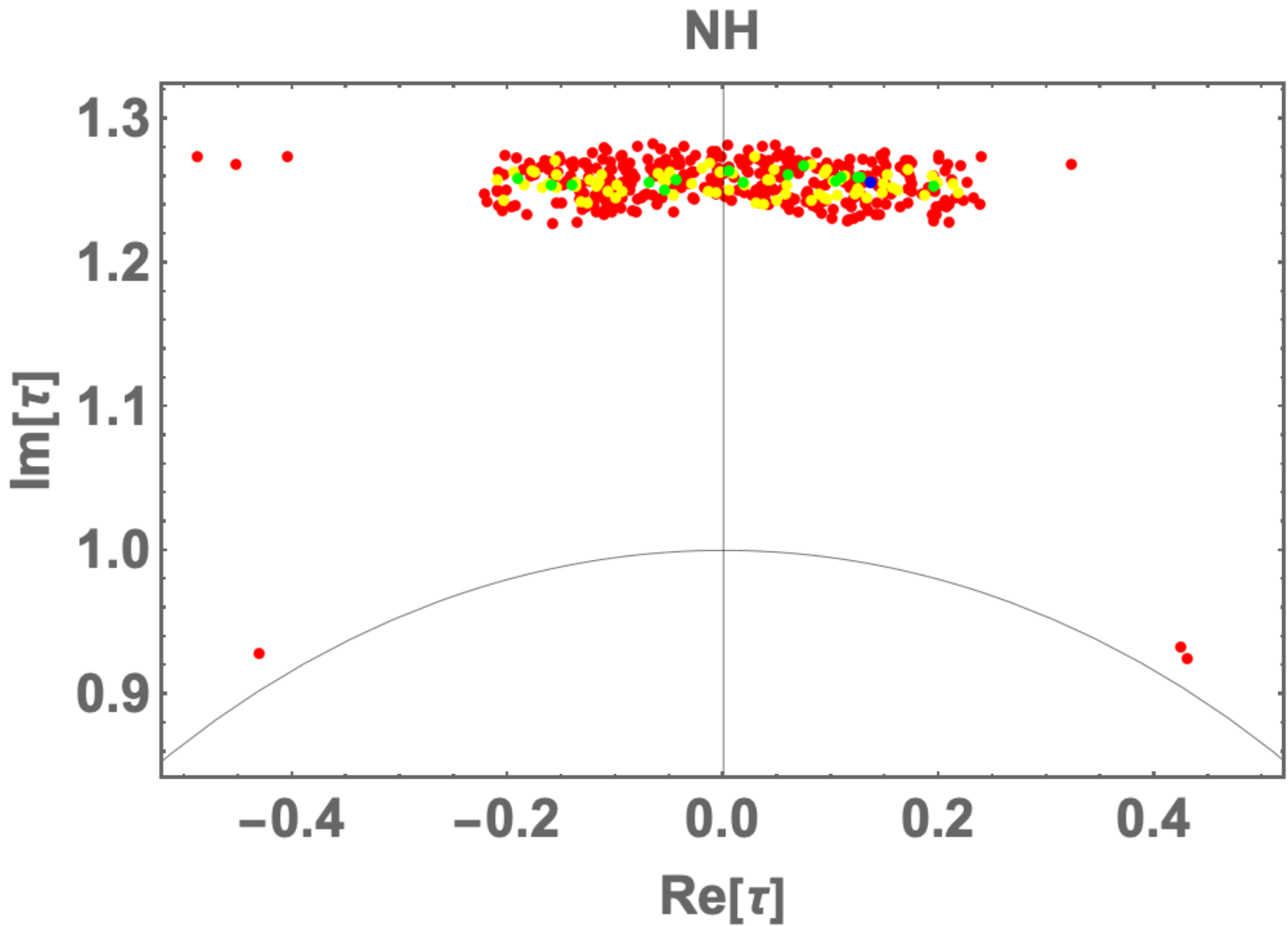}
  \caption{Allowed region for real $\tau$ and imaginary $\tau$ in NH.}
  \label{fig:tau_nh}
\end{figure}
{In Fig.~\ref{fig:tau_nh}, we show the allowed region of $\tau$, and find that the allowed region is concentrated at nearby $|{\rm Re}[\tau]|=[0.0-0.2]$ and ${\rm Im}[\tau]=[1.26-1.28]$ where the value is close to the fixed point $\tau =i$. We also find a few points near the fixed point $\tau = \omega$.}~\footnote{Note here that these points are not sufficiently close to the fixed points to investigate the mass matrices analytically  by expanding modular forms in terms of deviation from the fixed points. To achieve such analysis, the absolute distance from the fixed points should be within 0.05. }

\begin{figure}[t]
  \includegraphics[width=77mm]{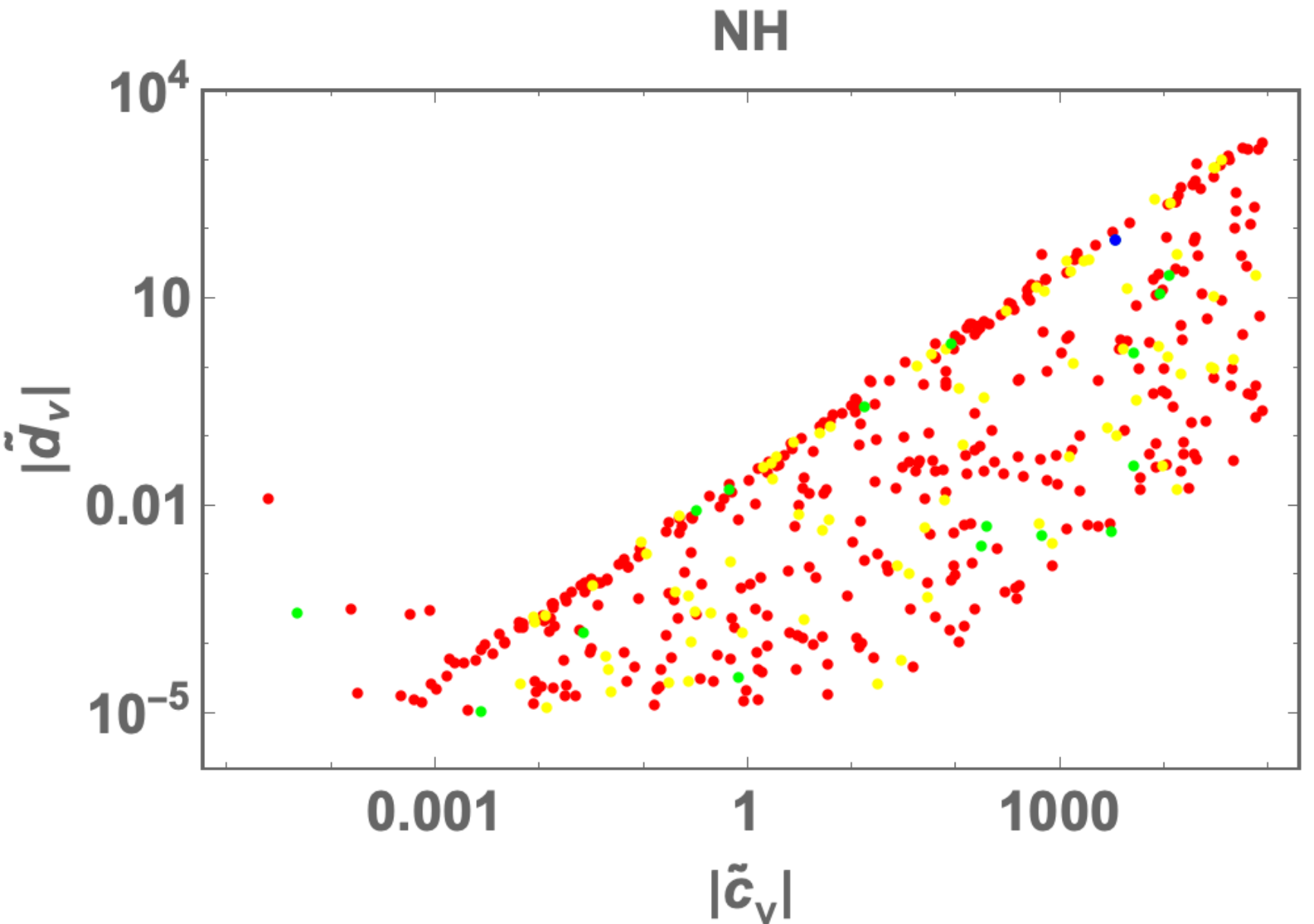}
  \includegraphics[width=77mm]{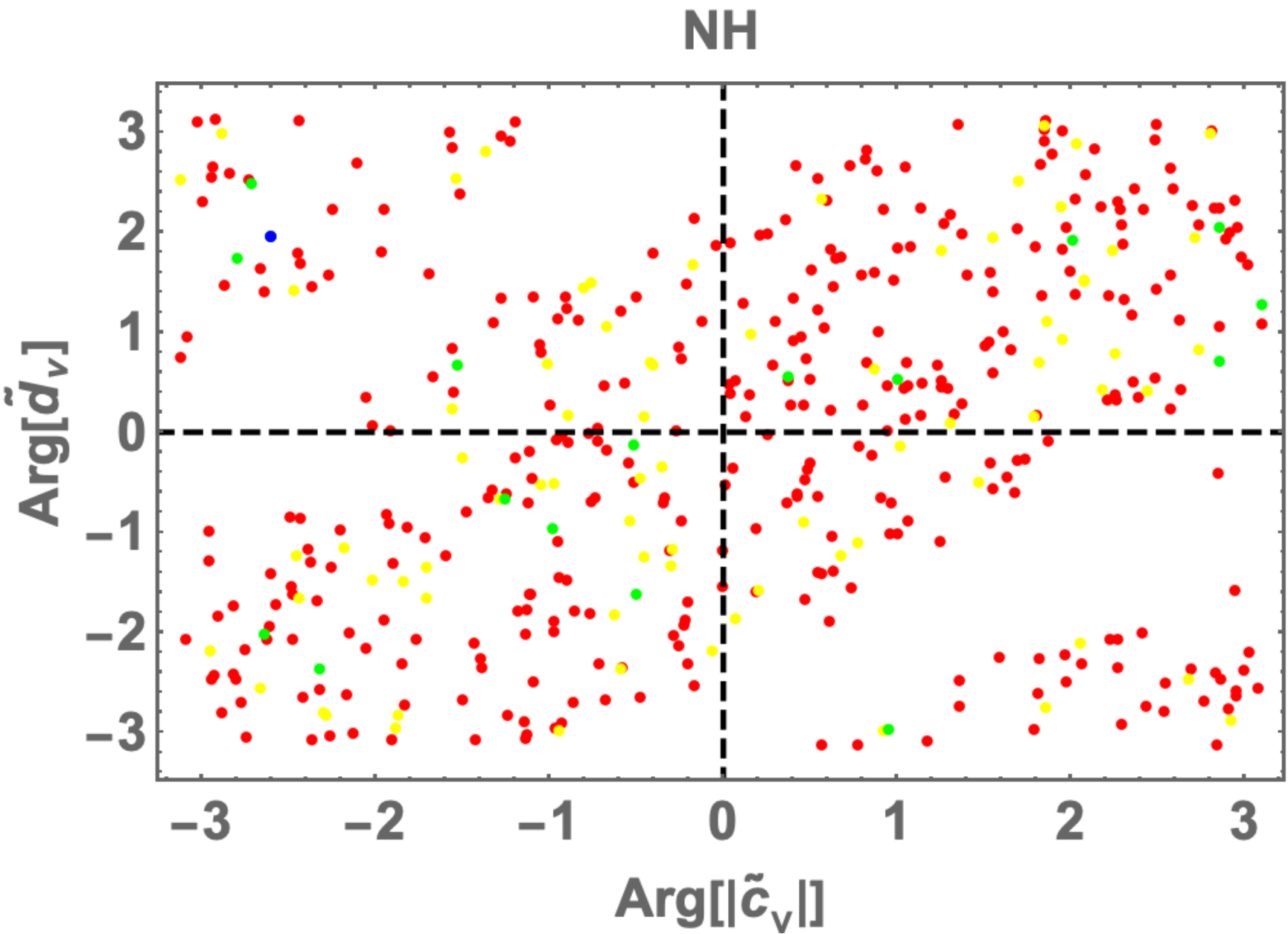}
  \caption{Allowed regions for absolute values (left) and argument ones (right) of $\tilde d_\nu$ and $\tilde c_\nu$ in NH.}
  \label{fig:cd_nh}
\end{figure}
In Fig.~\ref{fig:cd_nh}, we demonstrate the allowed regions for absolute values (left) and argument ones (right) of $\tilde d_\nu$ and $\tilde c_\nu$ in NH. We find that the allowed region is about
 $|\tilde c_\nu|=[10^{-4}-10^{5}]$ and $|\tilde d_\nu|=[10^{-5}-10^{4}]$ where $|\tilde{d}_\nu | \ll |\tilde{c}_\nu |$ is preferred ,
and
 ${\rm Arg}[\tilde c_\nu]$ and ${\rm Arg}[\tilde d_\nu]$ can be any value with little correlation.

\begin{figure}[t]
  \includegraphics[width=77mm]{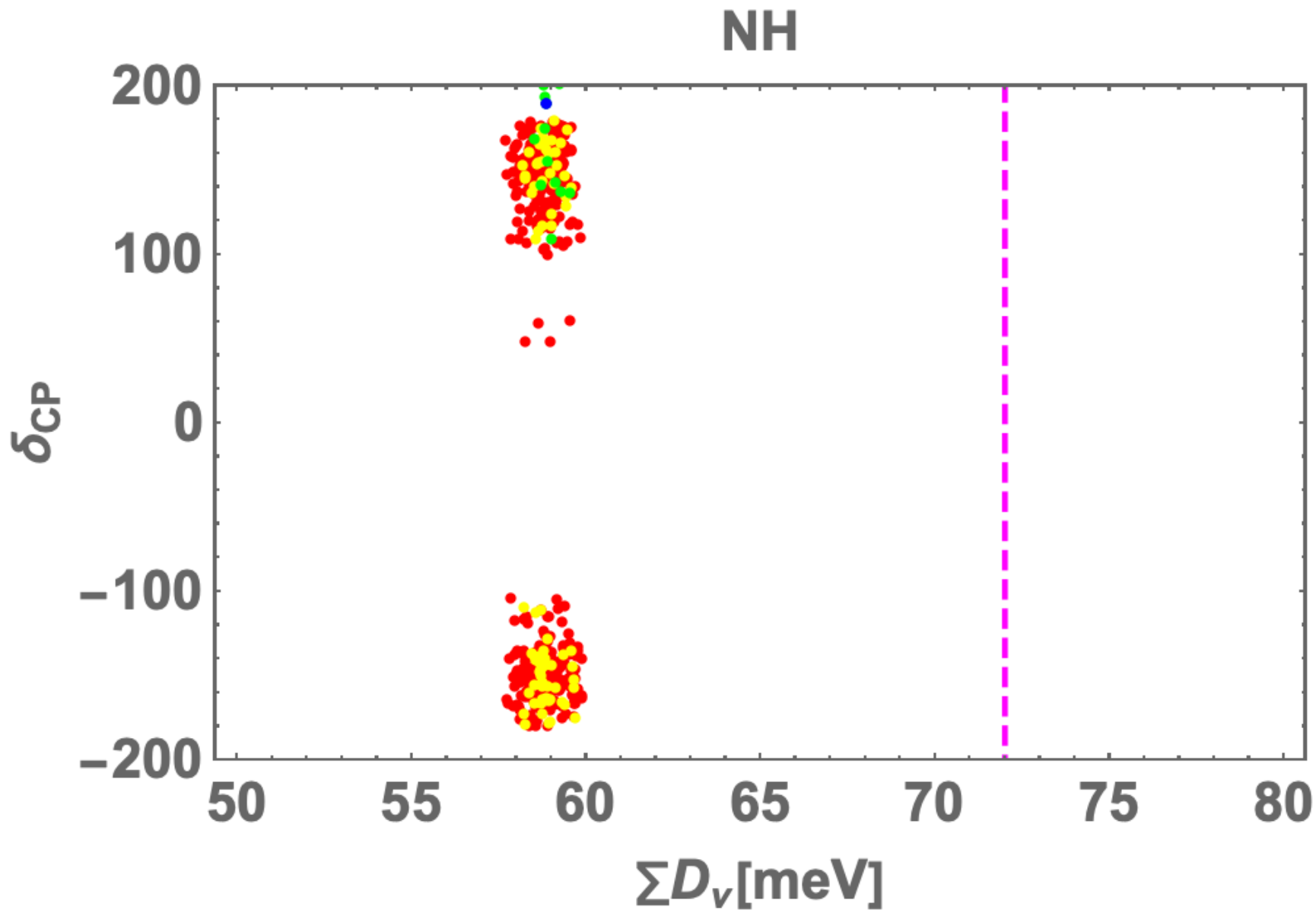}
    \includegraphics[width=77mm]{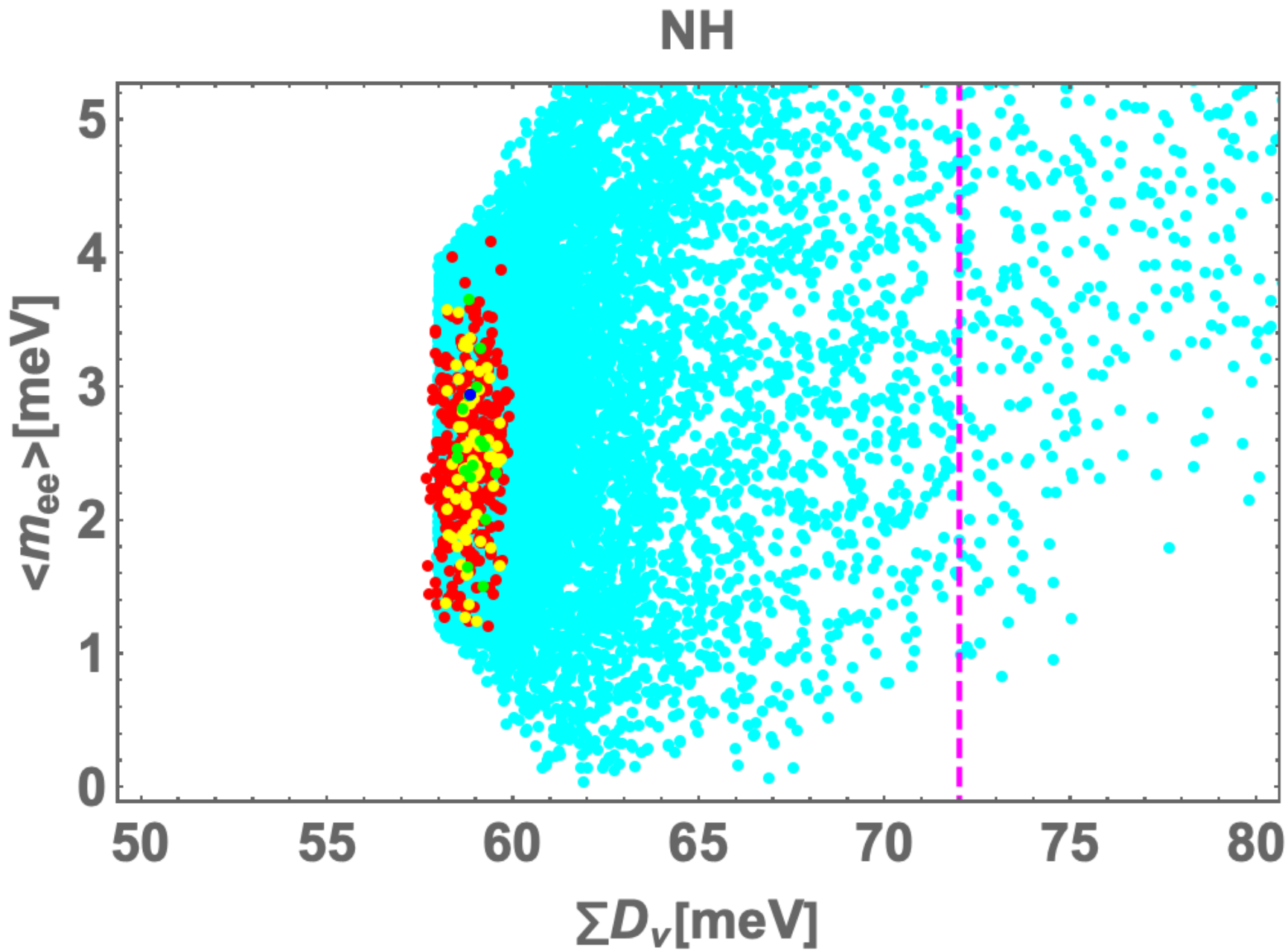}
  \caption{Allowed regions for  $\delta_{CP}$ deg (left) and $\langle m_{ee}\rangle$ meV (right) in terms of $\sum D_\nu$ meV in NH. The vertical magenta dotted line is upper bound on results of Planck+DESI~\cite{DESI:2024mwx} $\sum D_\nu\le $72 meV.   The cyan region in the left panel indicates allowed region by experimental result of Nufit 6.0.}
  \label{fig:dcp-mee_nh}
\end{figure}
In Fig.~\ref{fig:dcp-mee_nh}, we display the allowed  region for $\delta_{CP}$ deg (left) and $\langle m_{ee}\rangle$ meV (right) in terms of $\sum D_\nu$ meV. 
 We find the most of points are located at $|\delta_{CP}| = [90-200]$ deg and few points are around $\delta_{CP} = [40-60]$ deg. $\langle m_{ee}\rangle \approx [1-4]$ meV.
 The vertical magenta dotted line is upper bound on results of Planck+DESI~\cite{DESI:2024mwx} $\sum D_\nu\le $72 meV
 while $\sum D_\nu$ of our model is  $[58-60]$ meV which is nothing but a trivial consequence of two nonzero mass eigenvalues of active neutrinos.

\begin{figure}[t]
  \includegraphics[width=77mm]{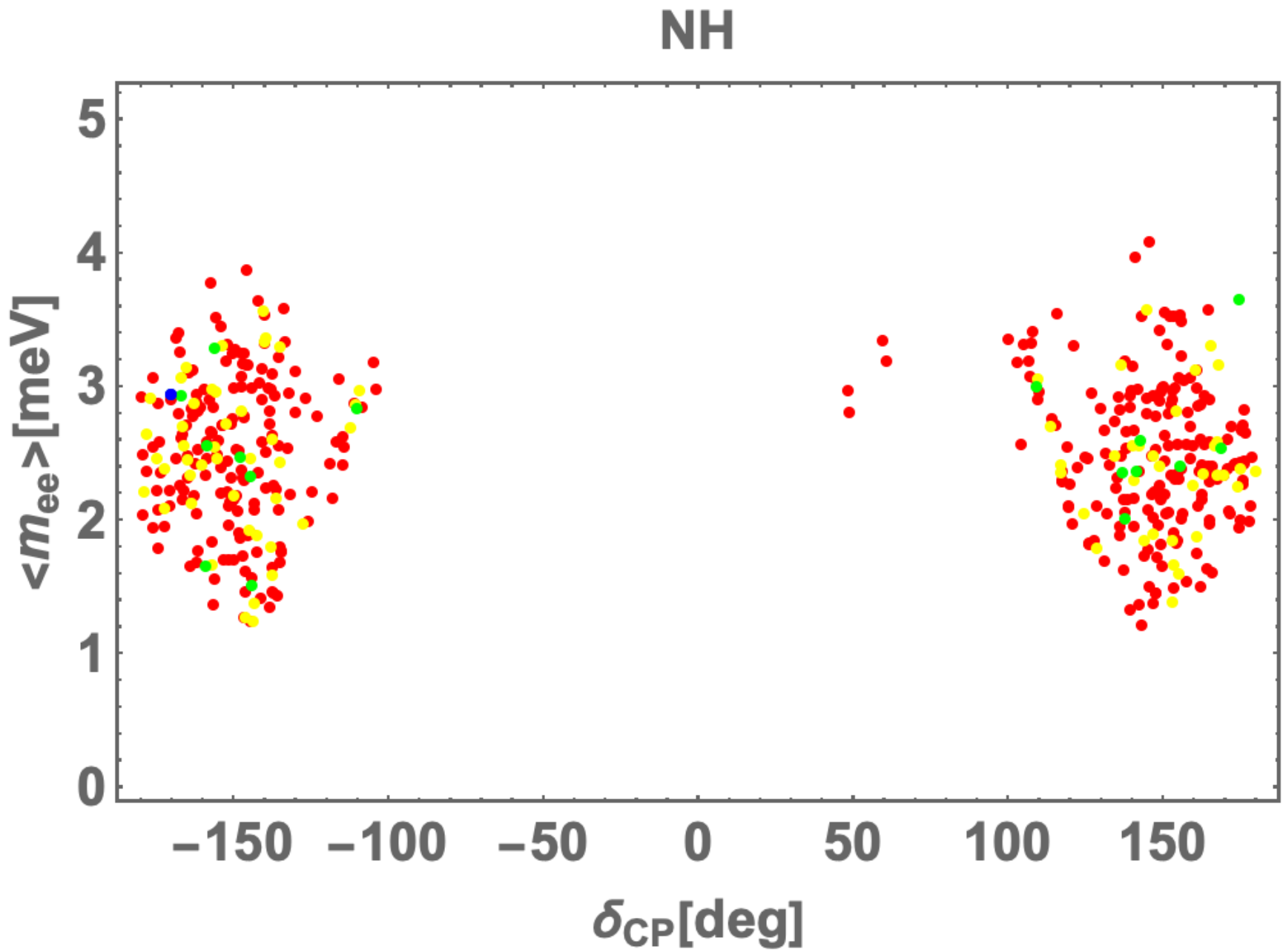}
    \includegraphics[width=77mm]{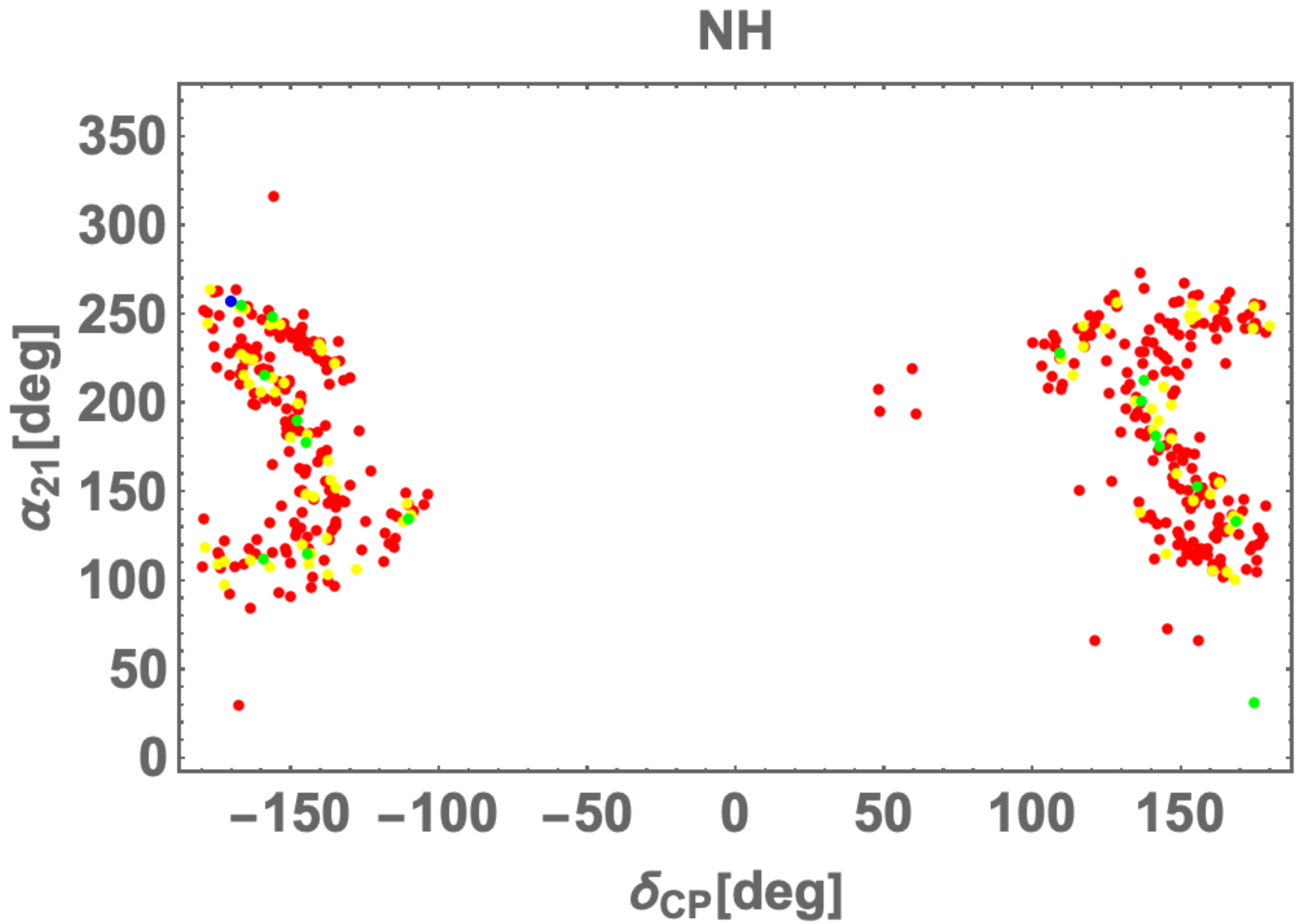}
  \caption{Allowed region for $\langle m_{ee}\rangle$ meV (left) and $\delta_{CP}$ deg (right) in terms of $\delta_{CP}$ deg in NH.}
  \label{fig:majo_nh}
\end{figure}
In Fig.~\ref{fig:majo_nh}, we show the allowed  region for $\langle m_{ee}\rangle$ meV (left) and $\alpha_{21}$ deg (right) in terms of $\delta_{CP}$ deg in NH. 
We 
find  the allowed region of $\alpha_{21}$ is  concentrated around $[80-270]$ deg with few points outside the region. 

We show a benchmark point (BP) that has the minimum $\Delta \chi^2$ in Table~\ref{bp-tab_i} and this BP will be employed to analyze the LFV, $g-2$, and DM in the next subsection.

\begin{table}[h]
	\centering
	\begin{tabular}{|c|c|c|} \hline 
			\rule[14pt]{0pt}{0pt}
 		&  NH  \\  \hline
			\rule[14pt]{0pt}{0pt}
		$\tau$ & $0.137 + 1.26  i$       \\ \hline
		\rule[14pt]{0pt}{0pt}
%
		$\tilde M_2$ & $5.34 \times 10^{-4}$   \\ \hline
		\rule[14pt]{0pt}{0pt}
		$\tilde c_\nu$ & $-2.85 \times 10^{3} - 1.69 \times 10^3 i$     \\ \hline
		\rule[14pt]{0pt}{0pt}
				$\tilde d_\nu$ & $-26.3 + 64.5 i$      \\ \hline
		\rule[14pt]{0pt}{0pt}
				$[a_e,a_\mu,a_\tau]$ & $[7.21 \times10^{-6}, -0.00139, 0.0206]$     \\ \hline
		\rule[14pt]{0pt}{0pt}
		$\Delta m^2_{\rm atm}$  &  $2.51 \times10^{-3} {\rm eV}^2$   \\ \hline
		\rule[14pt]{0pt}{0pt}
		$\Delta m^2_{\rm sol}$  &  $7.56 \times10^{-5} {\rm eV}^2$        \\ \hline
		\rule[14pt]{0pt}{0pt}
		$\sin\theta_{12}$ & $ 0.553$   \\ \hline
		\rule[14pt]{0pt}{0pt}
		$\sin\theta_{23}$ &  $ 0.683$   \\ \hline
		\rule[14pt]{0pt}{0pt}
		$\sin\theta_{13}$ &  $ 0.147$   \\ \hline
		\rule[14pt]{0pt}{0pt}
		$[\delta_{\rm CP}^\ell,\ \alpha_{21}]$ &  $[-170^\circ,\, 257^\circ]$   \\ \hline
		\rule[14pt]{0pt}{0pt}
		$\sum m_i$ &  $58.8$\,meV      \\ \hline
		\rule[14pt]{0pt}{0pt}
		$\langle m_{ee} \rangle$ &  $2.94$\,meV      \\ \hline
		\rule[14pt]{0pt}{0pt}
$\kappa$ &  $3.49\times 10^{-14}$      \\ \hline
		\rule[14pt]{0pt}{0pt}
		$\sqrt{\Delta\chi^2}$ &  $2.24$     \\ \hline
		\hline
	\end{tabular}
	\caption{Numerical {benchmark point (BP)} of our input parameters and observables in NH. Here,
	 this BP is taken  $\sqrt{\Delta \chi^2}$ to be minimum.
	 }
	\label{bp-tab_i}
\end{table}

\subsection{Numerical results of LFVs, lepton $g-2$, and DM in light of the neutrino results}

Before our numerical analysis, we prepare some definitions.
The neutrino mass matrix does not depend on all the masses inside the loop, but
the chi square analysis of the neutrino oscillation data gives us the value of $\kappa$.
While their masses inside the loop determine the values of LFVs, muon $g-2$, and the relic density of DM.
Thus, we rewrite Eq.~(\ref{eq:kappa}) as follows:
\begin{align}
\lambda_0= - \frac{(4\pi)^6}{(a_\nu b_\nu)^2} \left(\frac{\kappa M_1}{ m_\tau^2}\right).
\end{align}
When $a_\nu, b_\nu$, and $M_1$ are numerically fixed, $\lambda_0$ is numerically determined.
Then we impose the perturbative limit in our numerical analysis to be
\begin{align}
\lambda_0\lesssim \sqrt{4\pi}.
\end{align}

In addition, we restrict ourselves to be following conditions in order to forbid co-annihilation processes
and obtain the mass-independent loop function of the neutrino mass matrix:
\begin{align}
&1.2 m_\chi\le D_{N_2}\le D_{N_3},\\
&\epsilon_3 \le\frac15 ,\quad 0.9  m_{S_1}\le m_{S_2}\le  1.1 m_{S_1},
\end{align}
where we have defined $\epsilon_3$ to be $ \frac{D_{N_3}}{m_{S_1}}$.

Our input parameters randomly select within the following range:
\begin{align}
&\{ a_\nu, b_\nu \}  \in [0,\sqrt{4\pi}],
\quad%
M_1/{\rm GeV}   \in [10^{-5},10^5],
\end{align}
where  $a_\nu, b_\nu$ are real and the other needed parameters are employed by BP in the previous section.

In our numerical analysis, we found that Yukawa coupling $|b_\nu\times Y|$  exceeds the perturbative limit $\sim4\pi$ to obtain the observed relic density of DM while satisfying the constraints of LFVs and lepton universalities.
The correct relic density requires $ \mathcal{O}(100) \lesssim{\rm Max}[|b_\nu\times Y|]$ for NH case applying allowed parameters that can fit the neutrino data.
%
This implies that co-annihilations do not help reducing the Yukawa couplings to be perturbative limit. 
%
We could move to one of the next minimum model by changing the modular weight of $N_R$ to $-2$ instead of zero to obtain one more mass parameters giving wider region of allowed parameters, where the other assignments are exactly the same as our model.
However we would still have difficulty in realizing correct relic density while keeping perturbative limit for the Yukawa couplings. 
This is due to the fact that the DM annihilation cross section, Eq.~\eqref{eq:DMCX}, is p-wave dominant and we need relatively larger coupling constant than s-wave case.
In addition, neutrino data and LFV constraints require heavy DM and new scalars that  also suppress DM annihilation cross section.
It is thus difficult to obtain correct relic density in our minimal setting and some extension is necessary.


If we forget satisfying the observed relic density and we work our numerical analysis under the perturbative limit, 
we obtain the tendency for electron $g-2$, muon $g-2$, and LFVs as shown in Fig.~\ref{fig:lfvs_nh}.
These figures suggests that $-\Delta a_e$ and $-\Delta a_\mu$ are at most $10^{-20}$ and $10^{-15}$, respectively.
On the other hand, LFVs; especially $\mu\to e\gamma$ branching ratio, would be testable near future due to its maximum value is close to the experimental limit. 

\begin{figure}[t]
  \includegraphics[width=52mm]{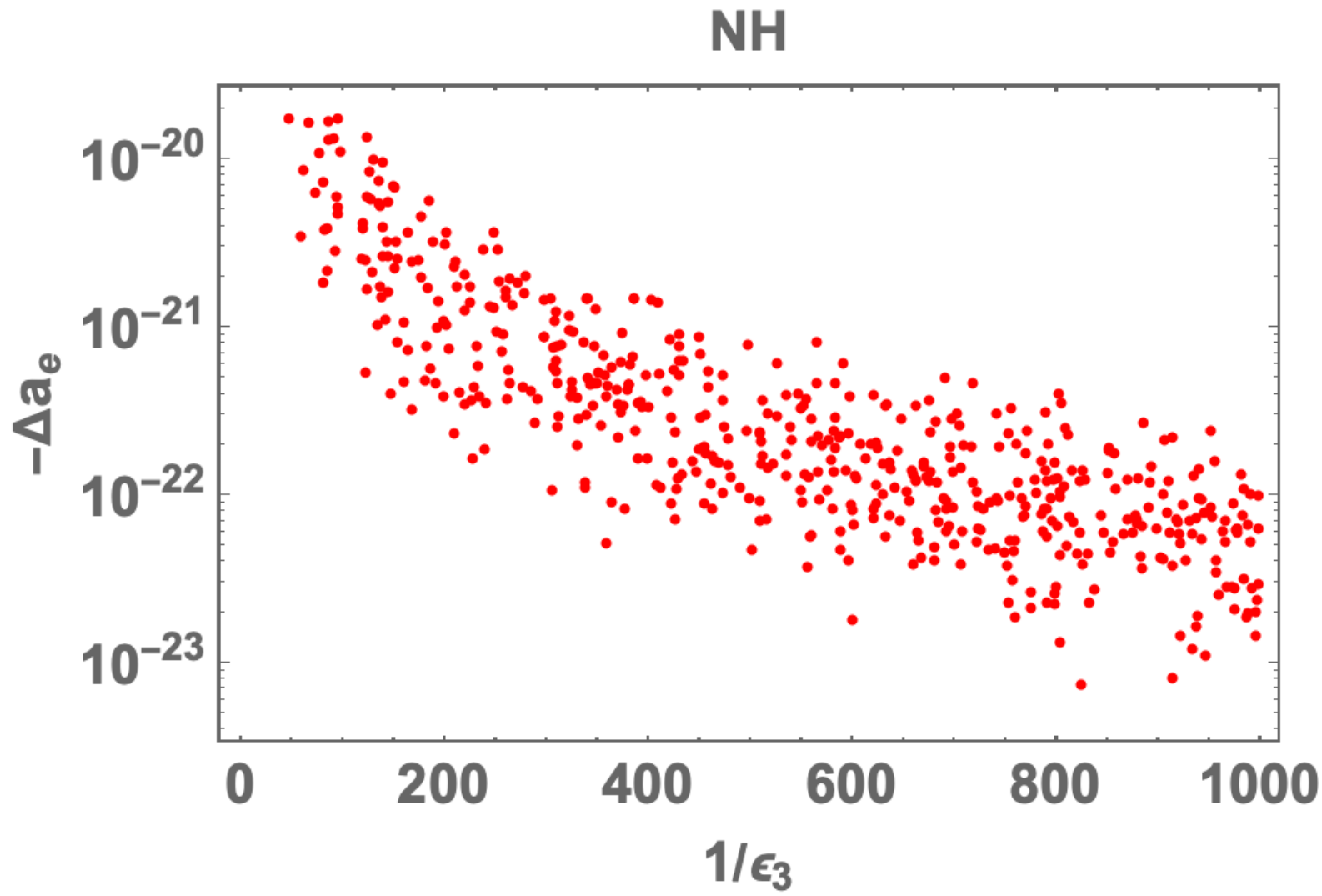}
  \includegraphics[width=56mm]{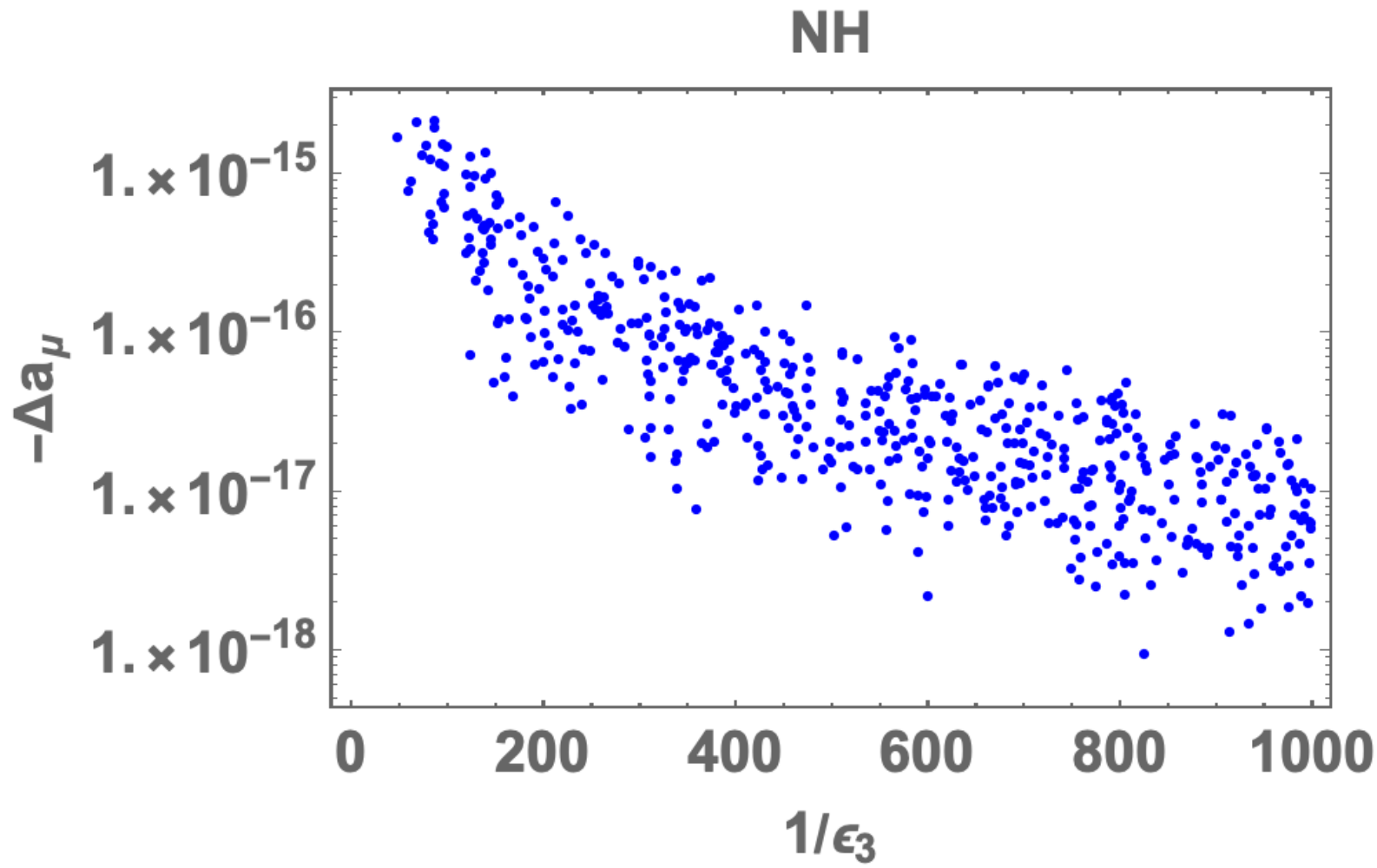}
    \includegraphics[width=52mm]{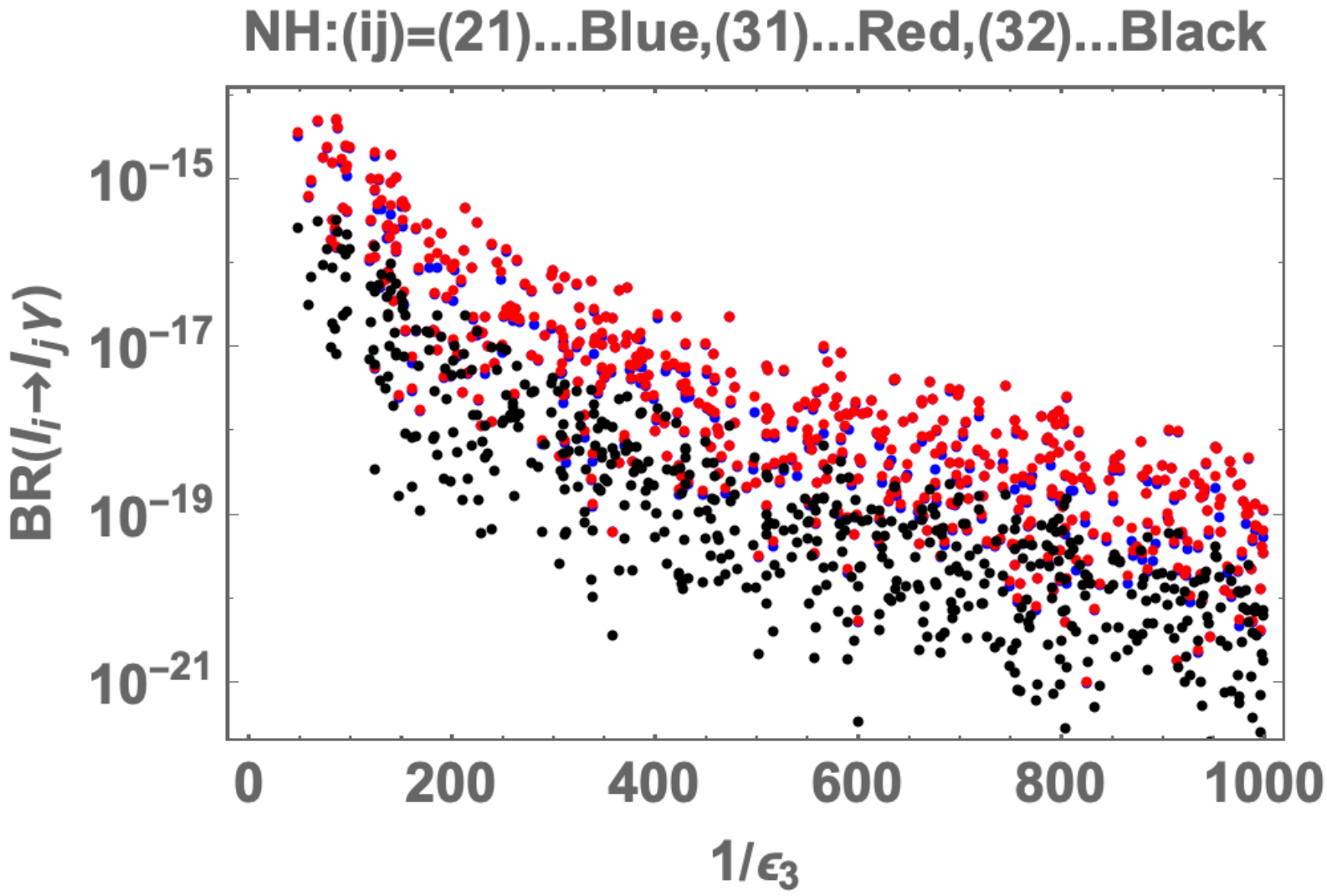}
  \caption{Allowed region for electron $g-2$(left), muon $g-2$(center), and LFVs(right), where these points do not satisfy the observed relic density.}
  \label{fig:lfvs_nh}
\end{figure}

\subsection{Minimal extension to accommodate relic density of DM}

We briefly illustrate one of the simplest solutions to explain the observed relic density without breaking our predictions for the neutrino sector, making use of a new interaction.
We can introduce a singlet scalar boson $S_0$ that leads to new interactions 
\begin{equation}
\mathcal{L}_{\rm new} = y_S S_0\overline {N_R^C}N_R  + \lambda_{\rm mix} S_0 H^\dagger H + \cdots,
\end{equation}
where its modular weight is {assigned to be zero for simplicity,}
assuming it is singlet under the $A_4$ symmetry, and we omit terms with $S_{1,2}^{\pm}$. 
We then have Higgs portal to the SM by mixing between $S^0$ and $h$ induced by the last term of $\mathcal{L}_{\rm new}$.
Note also that the addition of these interactions do not modify neutrino mass and the predictions in our analysis will not be changed.

As a result we have additional DM annihilation processes such as $\chi \chi \to S^0 \to f_{\rm SM} f_{\rm SM}$ and $\chi \chi \to S_0 S_0$.
In particular, s-channel cross section is useful to explain the relic density since annihilation cross section is enhanced at nearby $m_\chi\approx m_{S_0}/2$ where $m_{S_0}$ is the mass of $S_0$. 
The annihilation cross section of $\chi \chi \to S^0 \to f_{\rm SM} f_{\rm SM}$ process is approximately given by
\begin{equation}
(\sigma v_{\rm rel}) \simeq \frac{y_S^2 y_f^2 \sin^2 \alpha}{2 \pi} \frac{m_\chi^2}{(4 m^2_\chi - m^2_{S_0})^2},
\end{equation}
where $y_f$ is the SM Yukawa coupling for fermion $f$ and $\sin \alpha$ indicates the Higgs-$S_0$ mixing.
The relic density of DM is roughly estimated as $\Omega h^2 \sim 0.1 \ {\rm pb}/(\sigma v_{\rm rel})$ and we obtain 
\begin{equation}
\Omega h^2 \sim 0.12 \left( \frac{m_\chi}{1 \ {\rm TeV}} \right)^2 \frac{0.0081}{y_S^2 \sin^2 \alpha} \left( 1 - \frac{m_S^2}{4 m_\chi^2} \right)^{-2},
\end{equation}
where we considered top quark as $f$ for simplicity.
Thus we can realize $\Omega h^2 \sim 0.12$ with $m_\chi = 1$ TeV, $y_S =1$ and $\sin \alpha \sim 0.1 $ even if we don't have resonant enhancement.
With resonant effect, we can fit the relic density for small Higgs mixing case without conflicting constraints of direct detection searches~\cite{Kanemura:2010sh}.

\section{Conclusions and discussions}
We have investigated a three-loop induced neutrino mass model in a non-holomorphic modular flavor symmetry, in which we
found some prediction in a framework that masses inside the loop does not depend on structure of the neutrino mass matrix.
Since our model has a rank two Yukawa matrix in the neutrino sector, the lightest neutrino mass eigenvalue vanishes.
Here, we have realized a model with minimum free parameters;
three complexes $\tau,\ \tilde c_\nu,\ \tilde d_\nu$ and five reals $a_e,\ a_\mu,\ a_\tau,\ \tilde M_2,\ \kappa$, due to an appropriate charge-assignments under the modular symmetry.
Then, we have performed the chi square analyses considering the neutrino oscillation data,
and in particular, we have found rather narrow arrowed regions  for the NH case while we could not fit the data in IH case.
By adopting the best fit value for  NH, we have further analyzed the lepton flavor violation, muon $g-2$, lepton flavor universalities, and dark matter,
where we have neglected all the complicated processes such as co-annihilation interactions by controlling the related masses.  
Through the numerical analyses, we have found it is difficult to explain the observed relic density within the perturbative limit.
But, it is easy to resolve it by introducing a singlet boson without changing predictions in neutrino sector.

\section*{Acknowledgments}
\vspace{0.5cm}
T.~N. is supported by the Fundamental Research Funds for the Central Universities. 
 H.~O. is supported by Zhongyuan Talent (Talent Recruitment Series) Foreign Experts Project. 


\appendix
\section{Loop function}
\label{app}
The loop function at the three level is in general obtained only via numerical way.
But if some conditions are imposed, one can analytically integrate it out.
Here, we show the integration under the case of $D_{N_i}
 \ll m_{S_{1,2}}$ to  which we apply our model where $m_{S_1}^2=m_{S_2}^2 \pm \delta m_S^2$ with $\epsilon_S\equiv \frac{\delta m_S}{m_{S_2}}\ll1$.
One can expand the integration in terms of $\epsilon_i (\equiv D_{N_i}/m_{S_1})$ and $\epsilon_S$ as 
\begin{align}
F&\approx a_0 + a_1 \epsilon_i^2 + b_1 \epsilon_S^2 + {\cal O}(\epsilon^{4}_{i})
+ {\cal O}(\epsilon^{4}_{S}),\\
a_0&\approx \int[dx]_3\int[dx']_3\int[dx'']_3
\left[
\frac{1}{\frac{y''(y+z)}{(1-z)z}+\frac{z''(y'+z')}{(1-z')z'}}
 \right],\\
 a_1&\approx -\int[dx]_3\int[dx']_3\int[dx'']_3
\left[
\frac{x''}{\left( \frac{y''(y+z)}{(1-z)z}+\frac{z''(y'+z')}{(1-z')z'} \right)^2} 
 \right],\\
 b_1&\approx \int[dx]_3\int[dx']_3\int[dx'']_3
\left[
\frac{(-1+z)z(-1+z')z' (-y y'' z'+y y'' z'^2-y' z z''+ y' z^2 z'')}
{\left(-y y'' z' -y'' z z' +y y'' z'^2 +y'' z z'^2-y' z z''+ y' z^2 z'' -z z' z''+z^2 z' z''\right)^2} 
 \right],
 \end{align}

Here $a_0\approx 0.062$, $a_1\approx -2.92$, and $b_1\approx -0.0281$ and
$\int [dx]_3\equiv \int_0^1dx\int_0^{1-x}dy|_{z=1-x-y}$.

\bibliography{NonHoloMa4_knt.bib}

\end{document}